\documentclass[acmsmall,screen]{acmart}
\usepackage{hyperref}
\usepackage{pdfcomment}
\usepackage{enumitem}
\newlist{tightitem}{itemize}{1}
\setlist[tightitem]{
  label={},        
  leftmargin=0pt,
  itemsep=0pt,
  topsep=1pt,
  parsep=0pt,
  partopsep=0pt
}
\newcommand{\Oone}{%
  \pdftooltip{O1}{Professional; Positive sentiment}%
}

\newcommand{\Otwo}{%
  \pdftooltip{O2}{Professional; Neutral sentiment}%
}

\newcommand{\Othree}{%
  \pdftooltip{O3}{Professional; Positive sentiment}%
}

\newcommand{\Ofour}{%
  \pdftooltip{O4}{Professional; Negative sentiment}%
}

\newcommand{\Ofive}{%
  \pdftooltip{O5}{Professional; Negative sentiment}%
}

\newcommand{\Osix}{%
  \pdftooltip{O6}{Professional; Neutral sentiment}%
}

\newcommand{\Oseven}{%
  \pdftooltip{O7}{Professional; Positive sentiment}%
}

\newcommand{\Oeight}{%
  \pdftooltip{O8}{Vernacular; Positive sentiment}%
}

\newcommand{\Onine}{%
  \pdftooltip{O9}{Professional; Positive sentiment}%
}

\newcommand{\Oten}{%
  \pdftooltip{O10}{Vernacular; Neutral sentiment}%
}

\newcommand{\Oeleven}{%
  \pdftooltip{O11}{Vernacular; Positive sentiment}%
}

\newcommand{\Otwelve}{%
  \pdftooltip{O12}{Professional; Positive sentiment}%
}

\newcommand{\Othirteen}{%
  \pdftooltip{O13}{Professional; Neutral sentiment}%
}

\newcommand{\Vone}{%
  \pdftooltip{V1}{Professional; partially aligned; medium AI reliance}%
}

\newcommand{\Vtwo}{%
  \pdftooltip{V2}{Professional; aligned; low AI reliance}%
}

\newcommand{\Vthree}{%
  \pdftooltip{V3}{Professional; aligned; medium AI reliance}%
}

\newcommand{\Vfour}{%
  \pdftooltip{V4}{Professional; partially aligned; high AI reliance}%
}

\newcommand{\Vfive}{%
  \pdftooltip{V5}{Vernacular; not aligned; high AI reliance}%
}

\newcommand{\Vsix}{%
  \pdftooltip{V6}{Professional; aligned; low AI reliance}%
}

\newcommand{\Vseven}{%
  \pdftooltip{V7}{Vernacular; aligned; medium AI reliance}%
}

\AtBeginDocument{%
  }
\usepackage{fontawesome5}
\usepackage{graphicx}
\usepackage{xparse}
\usepackage{enumitem}
\usepackage{pifont}
\usepackage{xcolor}
\usepackage{subcaption}
\usepackage{cleveref}
\usepackage{wrapfig}
\usepackage{tabularx}
\usepackage[table]{xcolor}
\usepackage{booktabs}
\usepackage{geometry}
\usepackage{array}
\usepackage{graphicx}  
\usepackage{enumitem}
\usepackage{pifont}
\usepackage{fontawesome5} 
\usepackage[most]{tcolorbox}
\usepackage{caption}
\usepackage{tablefootnote}
\usepackage{threeparttable}
\NewDocumentCommand{\cSay}{ o m }{%
  \textit{``#2''}
}
\setcopyright{acmlicensed}
\copyrightyear{2025}
\acmYear{2025}
\acmDOI{XXXXXXX.XXXXXXX}
\acmConference[FSE'26]{The ACM International Conference on the Foundations of Software Engineering}{July 05--09,
  2026}{Montreal, Canada}
\acmISBN{978-1-4503-XXXX-X/2018/06}

\newcommand{\yihung}[1]{\textcolor{blue}{[yihung: #1]}}   
\newcommand{\tom}[1]{\textcolor{red}{[Tom: #1]}}    
\newcommand{\jim}[1]{\textcolor{purple}{[Jim: #1]}} 
\newcommand{\vicki}[1]{\textcolor{orange}{[Vicki: #1]}} 
\newcommand{\boyuan}[1]{\textcolor{cyan}{[Gary: #1]}} 
\newcommand{\sarina}[1]{\textcolor{brown}{[Gary: #1]}} 
\newcommand{\krystal}[1]{\textcolor{olive}{[Gary: #1]}} 

\newcommand{\includeAuthorComments}[1]
{
   \ifthenelse{\equal{#1}{0}}
   {
      \renewcommand{\yihung}[1]
      {
         {} 
      }
      \renewcommand{\tom}[1]
      {
         {} 
      }
      \renewcommand{\jim}[1]
      {
         {} 
      }
      \renewcommand{\vicki}[1]
      {
         {} 
      }
      \renewcommand{\boyuan}[1]
      {
         {} 
      }
      \renewcommand{\sarina}[1]
      {
         {} 
      }
      \renewcommand{\krystal}[1]
      {
         {} 
      }
   }{}
}

\includeAuthorComments{1}

\begin{document}

\title{Building Software by Rolling the Dice: A Qualitative Study of Vibe Coding}
\author{Yi-Hung Chou}
\email{yihungc1@uci.edu}
\orcid{0000-0001-8537-0200}
\affiliation{%
  \institution{University of California, Irvine}
  \city{Irvine}
  \state{California}
  \country{USA}
}

\author{Boyuan Jiang}
\email{boyuanj@uci.edu}
\orcid{0009-0009-3085-8861
}
\affiliation{%
  \institution{University of California, Irvine}
  \city{Irvine}
  \state{California}
  \country{USA}
}

\author{Yi Wen Chen}
\email{sarinachen@mobagel.com}
\orcid{0009-0009-7048-4032}
\affiliation{%
  \institution{MoBagel Inc.}
  \city{Irvine}
  \state{California}
  \country{USA}
}

\author{Mingyue Weng}
\email{krystal@calerie.com}
\orcid{0009-0003-9410-3448}
\affiliation{%
 \institution{CalerieLife}
 \city{Irvine}
  \state{California}
 \country{USA}}

\author{Victoria Jackson}
\email{v.jackson@soton.ac.uk}
\orcid{0000-0002-6326-931X}
\affiliation{%
  \institution{University of Southampton}
  \city{Southampton}
  \state{Hampshire}
  \country{United Kingdom}
}

\author{Thomas Zimmermann}
\email{tzimmer@uci.edu}
\orcid{0000-0003-4905-1469}
\affiliation{%
  \institution{University of California, Irvine}
  \city{Irvine}
  \state{California}
  \country{USA}
}

\author{James A. Jones}
\email{jajones@uci.edu}
\orcid{0000-0002-5359-7934}
\affiliation{%
  \institution{University of California, Irvine}
  \city{Irvine}
  \state{California}
  \country{USA}
}

\renewcommand{\shortauthors}{Chou et al.}


\begin{abstract}
Large language models (LLMs) are reshaping software engineering by enabling vibe coding—building software primarily through prompts rather than writing code. 
Although widely publicized as a productivity breakthrough, little is known about how practitioners actually define and engage in these practices. 
To shed some light on this emerging phenomenon, we conducted a grounded theory study of 20 vibe-coding videos, including 7 live-streamed coding sessions ($\approx$16 hours, 254 prompts) and 13 opinion videos ($\approx$5 hours), supported by additional analysis of activity durations and intents of prompts. 
Our findings reveal a spectrum of behaviors: some vibe coders rely almost entirely on AI without inspecting code, while others examine and adapt generated outputs.
Across approaches, all must contend with the stochastic nature of generation, with debugging and refinement described as “rolling the dice.” 
Further, divergent mental models, shaped by vibe coders’ expertise and reliance on AI, influence prompting strategies, evaluation practices, and levels of trust. 
These findings open new directions for research on the future of software engineering and point to practical opportunities for tool design and education.

\end{abstract}



\received{11 Sept 2025}
\received[accepted]{22 Dec 2025}

\maketitle

\section{Introduction}
\begin{quote}
\textit{``There’s a new kind of coding I call ‘vibe coding’, where you fully give in to the vibes, embrace exponentials, and forget that the code even exists.''} \\
— Andrej Karpathy (Feb 2025), \href{https://x.com/karpathy/status/1886192184808149383}{Tweet introducing the term ``vibe coding''}
\end{quote}

With LLMs now integrated into modern IDEs such as Cursor~\cite{cursorsoftware} and Windsurf~\cite{windsurf}, as well as online coding platforms like VZero~\cite{noauthor_v0_nodate} and Lovable~\cite{noauthor_lovable_nodate}, developers can move beyond auto-completion to delegating full code generation, orchestrating multi-file edits, and even authorizing AI agents to execute shell commands. In this context, Karpathy’s notion of \textit{vibe coding} captures a style of software development in which natural language prompts drive the process, with direct editing and review of the underlying code becoming optional. 

\newcounter{toolsctr}      
\newcounter{autctr}        
\newcounter{desctr}        
\newcounter{relctr}        

\renewcommand{\thetoolsctr}{{\small\faTools}{\raisebox{.5ex}{\tiny\arabic{toolsctr}}}}
\renewcommand{\theautctr}{{\small\faDice}{\raisebox{.5ex}{\tiny\arabic{autctr}}}}
\renewcommand{\thedesctr}{{\small\faPencil*}{\raisebox{.5ex}{\tiny\arabic{desctr}}}}
\renewcommand{\therelctr}{{\small\faIcon[regular]{brain}}{\raisebox{.5ex}{\tiny\arabic{relctr}}}}

\newcommand{\findingtools}[2]{%
  \refstepcounter{toolsctr}%
  \noindent\thetoolsctr\ \textbf{#2}\ \label{#1}%
}

\newcommand{\findingaut}[2]{%
  \refstepcounter{autctr}%
  \noindent\theautctr\ \textbf{#2}\ \label{#1}%
}

\newcommand{\findingdes}[2]{%
  \refstepcounter{desctr}%
  \noindent\thedesctr\ \textbf{#2}\ \label{#1}%
}

\newcommand{\findingrel}[2]{%
  \refstepcounter{relctr}%
  \noindent\therelctr\ \textbf{#2}\ \label{#1}%
}
\begin{figure}
    \centering
    \begin{tcolorbox}[
      colback=white,
      colframe=black,
      title=\scriptsize\textbf{Qualitative Findings Summary}, 
      coltitle=black,
      fonttitle=\bfseries,
      colbacktitle=gray!20,
      boxrule=0.8pt,
      sharp corners,
      before skip=6pt,
      after skip=6pt,
      left=3pt, right=3pt, top=3pt, bottom=3pt,
      boxsep=2pt
    ]

{\scriptsize

{\scshape\bfseries Entities}
\vspace{1pt}\hrule\vspace{6pt}

\hspace{6pt}{\bfseries\color{black!80} Vibe Coders} \par\vspace{3pt}
\begin{tightitem}
  \item \hspace{6pt}\ref{finding:vibe-meaning} The meaning and purposes of vibe coding varied across vibe coders
  \item \hspace{6pt}\ref{finding:vibe-trust} Vibe coders purposively, unwillingly, or selectively trust AI
\end{tightitem}

\vspace{6pt}\hrule\vspace{6pt}

\hspace{6pt}{\bfseries\color{black!80} Wrappers \& Tools} \par\vspace{3pt}
\begin{tightitem}
  \item \hspace{6pt}\ref{finding:wrappers} Vibe coders mapped specific wrappers to tasks, combining them in workflows
  \item \hspace{6pt}\ref{finding:opaque-behavior} Opaque tool behaviors hindered accurate mental model updates
\end{tightitem}

\vspace{6pt}\hrule\vspace{6pt}

\hspace{6pt}{\bfseries\color{black!80} Foundation Models} \par\vspace{3pt}
\begin{tightitem}
  \item \hspace{6pt}\ref{finding:models}  Similar to the selection of wrappers, vibe coders purposively selected or combined models to achieve their goal
  \item \hspace{6pt}\ref{finding:unpredictability} Vibe coder illustrated the unpredictability of FM’s behavior while relying on evaluation practices
that are themselves unreliable
\end{tightitem}

\vspace{6pt}\hrule\vspace{6pt}

\hspace{6pt}{\bfseries\color{black!80} Artifacts} \par\vspace{3pt}
\begin{tightitem}
  \item \hspace{6pt}\ref{finding:artifacts-understanding} Vibe coders formed mental model of artifacts through prior experience, interactions with AI models, outputs, or directly from code
  \item \hspace{6pt}\ref{finding:prompt-shaped-by-understanding} Vibe coders' mental model around artifacts may affect prompting strategies
\end{tightitem}

\vspace{6pt}\hrule\vspace{6pt}

\hspace{6pt}{\bfseries\color{black!80} Prompt \& Context} \par\vspace{3pt}
\begin{tightitem}
  \item \hspace{6pt}\ref{finding:prompting-strategies} Vibe coders developed and employed strategies to execute requests, to explore, or to understand the current situation
  \item \hspace{6pt}\ref{finding:context-manage}{ Vibe coders actively manage the context to avoid over-indexing and reduce token size}
  \item \hspace{6pt}\ref{finding:tracking-shared} Keeping track of what has been shared with models has emerged as challenges for vibe coders
\end{tightitem}

\vspace{6pt}
{\scshape\bfseries Process}
\vspace{1pt}\hrule\vspace{6pt}

\hspace{6pt}{\bfseries\color{black!80} Design \& Requirements} \par\vspace{3pt}
\begin{tightitem}
  \item \hspace{6pt}\ref{finding:reqs-upfront-emergent} While some vibe coders prepared detailed requirements, many formulated new requirements in response to outputs
  \item \hspace{6pt}\ref{finding:fixation} Vibe coders typically accepted or refined the first generated outputs rather than seeking alternatives, signaling design fixation
\end{tightitem}
\vspace{6pt}\hrule\vspace{6pt}

\hspace{6pt}{\bfseries\color{black!80} Implementation \& Debugging} \par\vspace{3pt}
\begin{tightitem}
  \item \hspace{6pt}\ref{finding:stochastic-debug} Debugging or feature refinement—expected deterministic—can feel stochastic, like \textit{``throwing a dice''}
  \item \hspace{6pt}\ref{finding:mitigations} Vibe coders used undo, version control, small changes, automated tests, and linters to rein in stochastic changes
  \item \hspace{6pt}\ref{finding:wait-frustration} Waiting for the next output frustrated many; some mitigated by multitasking or sending prompts in parallel
\end{tightitem}

\vspace{6pt}\hrule\vspace{6pt}

\hspace{6pt}{\bfseries\color{black!80} Evaluation} \par\vspace{3pt}
\begin{tightitem}
  \item \hspace{6pt}\ref{finding:evaluation-variation} Vibe coders perform evaluations on multiple levels: code, outputs, or responses. Some delegated the evaluation to AI
\end{tightitem}

} 
\end{tcolorbox}

\caption{Preview of the Qualitative Findings: Our findings show that vibe coders construct opportunistic tooling and workflows (\faTools), navigate AI stochastic autonomy (\faDice), improvise requirements and design (\faPencil*), and engage in diverse strategies shaped by their reliance on AI and mental model (\faBrain).}
\label{fig:summary_of_qualitative_findings}
\end{figure}

Since its introduction, vibe coding has been presented as a transformative shift in software development, with both start-ups and enterprises reporting productivity gains~\cite{goode_why_nodate, mehta_quarter_2025}. News showed that benefits extend not only to professional developers but also to \emph{vernacular programmers}~\cite{shaw_myths_2020, kang_tldr_2024}—individuals who, though not formally trained as programmers or software engineers, create and adapt software to meet their own needs, ~\cite{oconnorVibeCodingNew2025, gengexploringstudentaiinteractionsvibe2025}.
However, skeptics caution that code produced via vibe coding is not ready for production, citing risks of hidden bugs and the ongoing need for human oversight—for example, an AI coding agent once deleted a production database despite explicit instructions to freeze code~\cite{varanasi_vibe_nodate, replitDeleteDatabase2025}.
  
With its popularity, the term \emph{vibe coding} has also become an umbrella for varied interpretations.
While some developers describe vibe coding as synonymous with fully agentic coding—delegating most of the process to autonomous systems through prompts entirely—others frame it as software engineering supported by agents that still involves manual code editing and inspection. 
Our interest is motivated by both their move beyond earlier forms of AI-assisted programming—such as GitHub Copilot, where support was largely limited to localized code auto-completion~\cite{ziegler2024measuring,bird_taking_2023,liang_large-scale_2024}—and their claimed accessibility, which enables vernacular programmers to engage in software creation with coding knowledge optional~\cite{oconnorVibeCodingNew2025,smithAIVibeCoding2025}.
We argue that the emergence of these agentic practices within multi-tool coding environments from coders with diverse background warrants closer empirical investigation into how people actually define and work with them.

To this end, we analyzed publicly shared videos, including live-streamed sessions—an increasingly common data source in the SE research community that provides a naturalistic view of workflows and think-aloud narratives~\cite{learning-chou-2025, alaboudi_exploratory_2019}—and post-hoc opinion videos where practitioners discussed their experiences. We use the term vibe coders to distinguish those who primarily build software by prompting LLMs from developers who author and maintain code directly.

To study this emerging phenomenon, we employed Straussian Grounded Theory~\cite{strauss_basics_2003}. We then used purposive sampling to capture heterogeneity across contexts~\cite{baltes_sampling_2022}, and conducted iterative open and axial coding. Triangulation was carried out both across data sources (live streams and opinion videos) and through researcher triangulation, involving researchers with different perspectives (engineers, designers, product managers). In total, we analyzed seven live streams ($\approx$16 hours) and 13 opinion videos ($\approx$5hrs). The conceptual framework was refined through continuous discussions among four software engineering researchers and three industrial practitioners. To deepen our analysis, we conducted a focused examination of live-stream videos to explore how vibe coders’ mental models shaped their prompting and time use. We compared coders who demonstrated awareness of the codebase with those who relied primarily on LLMs, analyzing 254 prompts and 2,439 recorded activities to capture patterns in time distribution, and the frequency of prompts with similar intents (``rolling the dice’’). Integrating these findings with our qualitative insights allowed us to strengthen the emerging conceptual framework of vibe coding.

Our findings, summarized in Figure~\ref{fig:summary_of_qualitative_findings}, show that vibe coders: construct opportunistic tooling and workflows (\faTools) by combining wrappers and models; navigate AI’s stochastic autonomy (\faDice), where even tasks that appear deterministic—such as feature refinement or debugging—become probabilistic and are managed with safeguards like small edits, version control, and tests; improvise requirements and design (\faPencil*) in ways that resemble exploratory programming~\cite{beth_kery_exploring_2017}, refining ideas in response to outputs but often anchoring prematurely on early results; and adopt diverse prompting, context-management, and evaluation strategies (\faBrain) shaped by mental models that evolve continuously yet can be inaccurate. 
\section{Motivation \& Background}
The role of LLMs in software development is evolving rapidly. Their use has expanded from external tools such as ChatGPT, to embedded assistants like early Copilot, and most recently to agentic-powered tool such as Cursor—capable of executing commands, editing multiple files, and autonomously retrieving external context. With its development, the software engineering research community is actively working to understand both the challenges and the opportunities they create.

\subsection{AI for SE: Evolving Role of AI for Practitioners}

\subsubsection*{AI as Coding Assistant}  
With the introduction of tools such as Copilot and ChatGPT, developers began constructing prompts to generate localized edits in software. Empirical studies report shifts in programmer behavior—from writing code directly to directing and interpreting AI-generated output—alongside perceived productivity gains, reductions in repetitive toil~\cite{ziegler2024measuring, bird_taking_2023, liang_large-scale_2024}, and learning through tutorial-like responses~\cite{meemWhyChatGPTPractitioners2025}. At the same time, their use requires significant auditing effort~\cite{gordon_co-audit_2023} to address hallucinations and security risks, as well as skill in crafting prompts with sufficient contextual detail~\cite{kemellStillJustPersonal2025,davilaIndustryCaseStudy2024,weiszExaminingUseImpact2025}. AI functions primarily as an assistant—providing localized code suggestions and learning support while leaving overall direction, integration, and validation to the human developer.

\subsubsection*{AI as Programs}  
As practitioners increasingly began using prompts not only to query models but also to embed them as integral components of software systems, the research community has examined prompting as a first-class element of software~\cite{guy_prompts_2024,liang_prompts_2025,khojahPromptProgramming2025}. Liang et al.~\cite{liang_prompts_2025} highlight the challenges of integrating prompts into software systems given their stochastic and unpredictable behavior, while Khojah et al.~\cite{khojahPromptProgramming2025} report that although prompting strategies are abundant, their influence on function-level code generation quality is often limited. In other words, while integration of AI enables intelligent features, programmers still contend with non-determinism, relying on trial-and-error and extensive testing to ensure reliability.

\subsection{Vibe Coding: AI Roles as Co-author, Worker, or Beyond}  
Recently, the adoption of AI in software development has become mainstream in industry—with 84\% of respondents reporting usage by 2025 Stack Overflow survey~\cite{stackOverflowDeveloperSurvey2025}. Building on the rapid advances in foundation models (FMs) and supporting toolchains, such as autonomous agents and the Model Context Protocol, Andrej Karpathy introduced the term ``vibe coding''~\cite{andrejkarpathy} to describe instructing AI through natural language and having it carry out the software creation process on the developer’s behalf. In vibe coding, AI takes on a co-authoring role by generating larger units of functionality in response to high-level instructions. While humans continue to validate and integrate outputs, much of the lower-level coding labor is delegated to the AI. As a result, programmers—and even vernacular ones—have begun employing AI agents to author software on their behalf, with programming knowledge optional. Popular press highlights its democratizing potential for general public~\cite{oconnorVibeCodingNew2025} and its value for rapid prototyping~\cite{smithAIVibeCoding2025}, but also warns of risks including poor code quality, hallucinations, and security vulnerabilities~\cite{oconnorVibeCodingNew2025,rooseNotCoderAI2025,stokel-walkerWhatVibeCoding2025,williamsWhatVibeCoding2025,smithAIVibeCoding2025,replitDeleteDatabase2025}. Our work engages these simultaneous waves of hype and concern, contributing to ongoing debates over how AI will reshape the trajectory of software engineering.

\section{Research Method}
With vibe coding still an emerging phenomenon lacking established frameworks, we employed a Straussian Grounded Theory~\cite{strauss_basics_2003} to inductively develop insights into vibe coders’ practices and strategies. This allowed us to capture the diverse ways vibe coding was defined and enacted, and informed subsequent analyses examining how these practices shaped developers’ time allocation and prompting strategies.
\begin{figure}
    \centering
    \includegraphics[width=\linewidth]{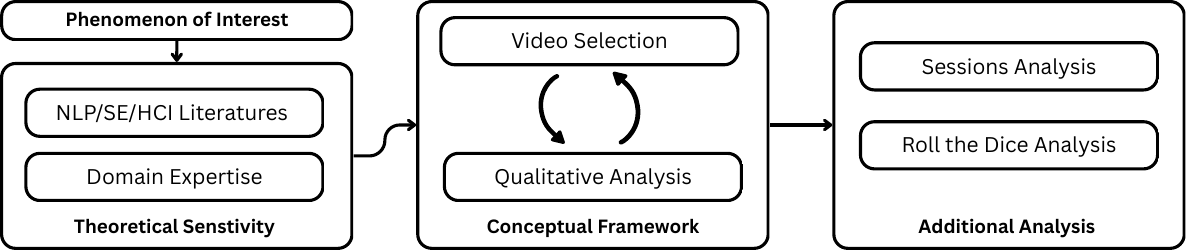}
    \caption{A summary of our Grounded Theory process. We first developed the theoretical sensitivity through literatures and domain expertise, then conduct iterative grounded theory development. Finally, we deepen the understanding of the conceptual framework with addition mix-method analysis on the live-stream videos.}
    \label{fig:resarch_process}
\end{figure}
\subsection{Grounded Theory Process}
Grounded Theory is a qualitative method that seeks to build conceptual understanding directly from data rather than test pre-existing theories. Through iterative analysis and constant comparison, researchers generate categories and relationships that can take the form of theories, models, or frameworks. In this study, we adopted the Straussian Grounded Theory approach~\cite{strauss_basics_2003} for understanding vibe coding practices. We depicted the process in Figure \ref{fig:resarch_process}, which contains four main stages: identifying a phenomenon of interest, developing theoretical sensitivity, generating conceptual framework, deepening understanding through additional analysis. 

\begin{table}[t]
\centering
\begin{threeparttable}
  \caption{Live-stream videos meeting selection criteria.\label{Livestream_Video_Subjects}}
  \tiny
  \renewcommand{\arraystretch}{1.2}
  \begin{tabular}{ccp{2.5cm}ccp{2.5cm}p{2.2cm}p{1.3cm}}
    \toprule
    \textbf{ID} & \textbf{Time} & \textbf{Problem Domain} & \textbf{\# Activities} & \textbf{\# Prompts} & \textbf{Tools} & \textbf{Vernacular/Professional} & \textbf{Alignment}\tnote{a} \\
    \midrule
   V1   & 1:53:00 & Web (Python, JavaScript)        & 360 & 38 & Cursor                         & Professional & Partially \\
    V2   & 2:20:00 & Discord Bot (JavaScript)        & 117 & 14 & ChatGPT, VSCode                & Professional & Align \\
    V3\tnote{b} & 25:44   & Web (Ruby)                       & 75  & 10 & ChatGPT, Cursor                & Professional & Align \\
    V4\tnote{b} & 28:21   & Game (JavaScript)               & 137 & 38 & Cursor                         & Professional & Partially \\
    V5  & 4:57:56 & Web (Python, JavaScript)        & 923 & 79 & Claude Code, ChatGPT           & Vernacular   & Not Align \\
    V6   & 4:22:00 & Web (JavaScript)                & 661 & 54 & Windsurf, ChatGPT, VZero       & Professional & Align \\
    V7   & 1:04:35 & Mobile App (TypeScript)         & 166 & 21 & Cursor                         & Vernacular   & Align \\
    \bottomrule
  \end{tabular}
  \begin{tablenotes}
    \item[a] “Alignment” indicates whether the streamer’s prior technical background aligns with the project they are working on during the live stream.
    \item[b] V3 and V4 are edited videos; consequently, the frequency of switching activities is higher than in the other videos.
  \end{tablenotes}
\end{threeparttable}
\end{table}
\begin{table}[t]
\centering
\begin{threeparttable}
  \caption{Summary of Opinion Vibe Coding Videos.\label{tab:vibe_coding_background}}
  \tiny
  \begin{tabular}{ccclc}
    \toprule
    \textbf{ID} & \textbf{Time} & \textbf{Sentiment} & \textbf{Tools} & \textbf{Vernacular/Professional Programmers} \\
    \midrule
    O1   & 27:16          & Positive  & Claude Code / Cursor        & Professional  \\
    O2   & 24:54          & Neutral   & Not Specified               & Professional  \\
    O3   & 21:38          & Positive  & Not Specified               & Professional  \\
    O4\tnote{a} & 55:09          & Negative  & Not Specified               & Professional  \\
    O5\tnote{a} & 55:09          & Negative  & Not Specified               & Professional  \\
    O6   & 24:52          & Neutral   & Claude Code, Cursor         & Professional  \\
    O7   & 16:39          & Positive  & Claude Code, Windsurf, Aqua & Professional  \\
    O8\tnote{b} & 11:08 + 10:36  & Positive  & Bolt, ChatGPT               & Vernacular \\
    O9   & 33:22          & Positive  & Replit, Cursor            & Professional  \\
    O10  & 47:29          & Neutral   & Cursor, Claude Code, VZero     & Vernacular \\
    O11  & 14:32          & Positive  & Vzero                       & Vernacular \\
    O12  & 10:10          & Positive  & VSCode                      & Professional \\
    O13  & 7:09           & Neutral   & Not Specified               & Professional \\
    \bottomrule
  \end{tabular}
  \begin{tablenotes}
    \item[a] O4 and O5 appear in the same video, with O5 providing reflections while viewing O4.
    \item[b] Two videos are included from O8, documenting experiences after 30 and 60 days of vibe coding.
  \end{tablenotes}
\end{threeparttable}
\end{table}
\subsubsection{Identifying a Phenomenon of Interest}
We began with an open-ended question: ``How do vibe coders define and engage in vibe coding'' At this stage, the research focus was intentionally broad. To orient the inquiry, we examined the phenomenon of individuals using AI-powered development tools to construct software primarily through prompting.

\subsubsection{Developing Theoretical Sensitivity}
The process of selecting codes and establishing relationships among them requires authors to exercise \textit{theoretical sensitivity}~\cite{stol_grounded_2016}, that is, the capacity to discern what is meaningful in the data. 
Some of the contributors to this work were invited as industry collaborators, thus did not have prior exposure to relevant literature.
As such, 
before engaging with the dataset, the academic authors assigned a collection of relevant literature and documentation. 
These materials include prior relevant literatures~\cite{liang_prompts_2025, barke_grounded_2023, zamfirescu-pereira_why_2023}, and documentation for contemporary AI-powered coding tools~\cite{noauthor_introducing_nodate, v0docs}. 
The purpose of this preparatory activity is to establish a shared vocabulary and a common conceptual foundation among the authors. To further align perspectives, the first author also facilitated an orientation session that debriefed methodological resources, drawing on key materials such as Stol et al.~\cite{stol_grounded_2016} and Gibbs~\cite{noauthor_grounded_nodate}.

\subsubsection{Generating Conceptual Frameworks\label{conceptual_framework}}
Our process followed the principles of grounded theory and unfolded through two tightly connected activities: \textbf{video selection} and \textbf{qualitative analysis}. These activities were conducted iteratively and in parallel, with insights from analysis shaping subsequent sampling decisions, and newly sampled videos in turn enriching the analytic process.
\paragraph{Video Selection} We constructed the dataset in two intertwined steps—search-based selection and purposive sampling. First, we curated two complementary sources: (1) livestream or minimally edited livestream videos to observe real-time prompt evolution, evaluation strategies, and development workflows, and (2) opinion-oriented videos in which developers offered post-hoc advice and reflections on \textit{vibe coding}. To reduce algorithmic bias, each author searched YouTube with the identical keyword “Vibe Coding,” excluded YouTube Shorts, and sequentially logged results in a shared Google Sheet to ensure transparency and prevent duplication. Because long-form livestreams were scarce, some authors supplemented queries with “Live” and filtered for videos longer than 20 minutes. The initial search yielded 40 videos. After discussion and unanimous voting, we excluded advertisements and non-English content, resulting in 3 livestreamed videos, 2 near-livestreamed videos, and 6 opinion-based videos. Guided by purposive sampling with maximum variation~\cite{baltes_sampling_2022, suri_purposeful_2011}—seeking diversity in technical backgrounds, alignment between background and task domain, problem types, and tools—and informed by early analytic gaps, we ran a second targeted search using “no experience” and “0 experience,” yielding 20 unique videos, from which we selected 7  opinion-based and 2 livestreamed videos to incorporate novice perspectives. Table~\ref{Livestream_Video_Subjects} summarizes livestreamers’ tasks, total activities, and prompts, and indicates whether coders were vernacular or professional programmers and whether their technical backgrounds aligned with the problem domain. Background information came from self-disclosures during videos (e.g., \Vsix{} noting, \cSay[V6]{We may have worked at Facebook}) and publicly available sources (e.g., LinkedIn profiles, personal websites). Alignment was determined by cross-referencing disclosed backgrounds with the domains of tasks undertaken (e.g., \Vone, a data scientist, worked on a web-server task outside his usual expertise, albeit still in Python). In parallel, Table \ref{tab:vibe_coding_background} provides an overview of the opinion-oriented videos, highlighting sentiments, tools, and programmer backgrounds.
\paragraph{Qualitative Analysis} The first four authors independently analyzed the three initial livestream videos (V1–V3), applying open and axial coding to develop preliminary insights. Early analysis highlighted the importance of ad-hoc expressions of frustration and excitement—yet these moments were much rarer in livestreams than in opinion videos, where coders explicitly reflected on such experiences. Because such high-signal moments were scarce and livestream content itself was limited, we incorporated opinion-oriented videos, whose post-hoc reflections complemented the livestream data and enriched the theoretical development. The remaining videos were coded independently using open coding. To ensure consistency, authors met weekly to review and compare new codes against existing ones; naming, inclusion, or exclusion required unanimous agreement. Disagreements were resolved through discussion; for example, in early coding process, a mistake by a vibe coder was labeled “unintentional error” by one author and “tool constraint” by another. Rather than collapsing these differences, both codes were retained to capture the range of valid interpretations. These analytic insights fed back into subsequent video selection. We observed theoretical saturation after analyzing V5 and O8: no new codes or categories emerged, and subsequent videos reinforced rather than extended the existing framework.

\subsection{Threats to Validity}
Our methodological approach follows Creswell and Poth~\cite{creswell_qualitative_2016}, consistent with prior SE research~\cite{jackson_co-creation_2024}. We employed triangulation reflexivity, external auditing, and peer debriefing to ensure credibility, reliability, and transferability.\paragraph{Credibility.}
We employed \textbf{investigator triangulation} to reduce the influence of individual bias. Our coding team included one developer, one SE researcher, one product designer, and one product manager. Each author independently created an initial codebook based on the first three live-stream videos (\Vone, \Vtwo, \Vthree) to maximize variation. Differences in coding—shaped by diverse professional backgrounds and attitudes toward vibe coding—were explicitly discussed.
Following de Sousa’s guidance on \textbf{reflexivity}~\cite{de_sousa_integrating_2025}, and before beginning data collection, each author drafted reflective notes to articulate their preconceptions and positionalities. These notes captured the range of perspectives within the team—optimism from the product designer and product manager, contrasted with skepticism from the software engineering researcher and developer—regarding the value and feasibility of vibe coding. By making these assumptions explicit initially, we acknowledged the potential biases that could shape our interpretations.
Throughout the study, we revisited these reflexive memos to examine how our perspectives influenced both the coding process and the emerging theory. For example, the researchers with programming experience might be inclined to attribute user mistakes to carelessness, whereas the designer might instead view them as stemming from shortcomings in the tool’s design. This iterative reflexive practice helped ensure that our interpretations remained critically self-aware rather than unconsciously reproducing our biases.

\paragraph{Reliability} We shared our study design and sampling method in §\ref{conceptual_framework} and made the codebook available at~\cite{Dryad:2jm63xt31}.
In addition, we engaged in \textbf{peer debriefing}: our team discussed findings regularly with two senior SE researchers who were not directly involved in the coding process, allowing them to challenge our assumptions and interpretations. A third researcher, acted as an \textbf{external auditor} by reviewing and discussing the coding process itself. These reviews provided an additional check on the dependability of our analysis and reduced the risk of unchecked bias.

\paragraph{Transferability}
While all of our data was collected from YouTube, we aimed to cover a wide range of vibe coders through maximum variation sampling from different backgrounds (e.g., vernacular and professional programmers), formats (e.g., live-streamed and opinion-based), and tools (e.g., local code editors such as Cursor and online no-code platforms such as VZero). Some video creators could be identified as based in the UK or USA, but for most we could not determine detailed demographic. The nature of live-stream video might also affect vibe coders' behaviors. For example, \Vone, when writing requirements, inclined to use AI as \textit{``we are just full out Vibe coding.''}, or \Vsix{} \textit{``let the vibe begins''} when seeing correct outputs.
Such experimental or exaggerated expressions highlight the performative dimension of live coding, which may limit the transferability of these findings to industry contexts, where legacy code, governance, and long-term maintainability impose stricter constraints.
Finally, the reliance on English-language videos enabled confident interpretation but may have constrained cultural diversity. Our dataset is also dominated by Python and JavaScript, likely reflecting both tool affordances and language popularity. Future research should extend beyond English and across a broader range of programming languages to test whether these findings generalize across cultural and technical contexts.
\subsubsection{Ethical Considerations}
This study followed our institution’s ethics guidelines. Because the dataset consisted only of publicly available, searchable YouTube videos, formal IRB approval was not required. YouTube’s Fair Use policy~\cite{youtubeFairUse} permits the legal use of such material for research, and internet research ethics guidance~\cite{townsend2016social,hoda_socio-technical_2022} recommends anonymization of public data to reduce potential harm, which we applied in our analysis.

\begin{figure}
    \centering
    \includegraphics[width=\linewidth]{Figures/model.pdf}
    \caption{\textbf{Vibe coding across levels of AI reliance.} Coders draw on prior experience to form an initial mental model and provide prompts to the Wrapper and Foundational Model (FM). In high-reliance mode, they primarily evaluate artifacts via model responses and system outputs (e.g., UI displays or logs). In low-reliance mode, they may also edit code directly and assess both implementation and outputs. Evaluation in either mode serves to refine their mental model of the entities involved.}
    \label{fig:vibe_coding_general}
\end{figure}
\section{Results}

After axial coding, we identified five entities in the vibe coding process: Vibe Coders, vernacular or professional programmers who carried out the work (§\ref{vibe_coders}); Prompts \& Contexts, the inputs and supporting information they provided (§\ref{prompts}); Wrappers \& Tools, the interfaces through which coders interacted with foundational models (§\ref{wrappers}); the Foundation Model (FM), the underlying AI models that generated responses and code (§\ref{FM}); and Artifacts, the outputs of the models, including generated code and system outputs (§\ref{artifacts}). 
Vibe coders formed mental models\footnote{In this context, we use Norman's definition \cite{norman_observations_1983} of a \emph{mental model}, which is simply a person's  internal representation ``of themselves and of the things with which they are interacting\ldots [to] provide predictive and explanatory power for understanding the interaction.''} around the above mentioned entities to select appropriate wrappers, tools, and FMs, and to compose prompts~\cite{latoza_maintaining_2006, norman_observations_1983}, through interacting with artifacts and through model responses, the mental model get updated.

Figure~\ref{fig:vibe_coding_general} 
illustrates the overall process that we observed. Vibe coders first sent the prompt. Once the FM processed their requests, vibe coders either directly edited the generated code or files, or issued additional prompts to debug or introduce new features. These iterative cycles comprised three higher-level activities: Requirements \& Design (§\ref{requirement}), Implementation \& Debugging (§\ref{implementation}), and Evaluation (§\ref{evaluation}). Below, we examine each entity and activity in detail.

\subsection{Vibe Coders\label{vibe_coders}}
\subsubsection*{\findingrel{finding:vibe-meaning}{The meaning and purposes of vibe coding varied across vibe coders.}} While vibe coders did not always articulate explicit definitions, their behaviors often revealed an implicit understanding of the concept. We inferred these interpretations from observable actions—for example, \Vfive{} never opened a file across nearly five hours of recorded sessions. Table~\ref{tab:different_behavior} summarizes the key behaviors that reflect coders’ interpretations of vibe coding. The column labels are derived from the original vibe coding tweet (e.g., \textit{``I ‘Accept All’ always''} → Always Accept column), illustrating differences in how coders approached vibe coding. We also introduce an additional column, AI Reliance, constructed using the extraction criteria detailed in the replication package~\cite{Dryad:2jm63xt31}.
\begin{table}[t]
\centering
\begin{threeparttable}
\caption{Different Behaviors when Vibe Coding for Live-Streamed Coders\label{tab:different_behavior}}
\scriptsize 
\renewcommand{\arraystretch}{1.2}
\begin{tabular}{clllll}
\toprule
\textbf{ID} & \textbf{Input Modality} & \textbf{Accept All?} & \textbf{Reads Diffs or Code} & \textbf{Direct Copy of Error Messages\tnote{*}}  & \textbf{AI Reliance\tnote{\textdagger}} \\
\midrule
\Vone{} & Text & 1 Rejection & Yes & 9 / 41 & Medium \\ 
\Vtwo{} & Text, Image & Always Accept & Yes & None (Did not encounter errors) & Low \\ 
\Vthree{} & Text & Always Accept & Yes & 3 / 10 & Medium \\ 
\Vfour{} & Voice, Image & Always Accept & Never & 1 / 38 & High \\ 
\Vfive{} & Text & Always Accept & Never & 43 / 79  & High \\ 
\Vsix{} & Text, Image & 5 Rejections & Yes & 5 / 54 & Low \\ 
\Vseven{} & Text, Image & Always Accept & Never & 5 / 21 & Medium \\ 
\bottomrule
\end{tabular}
\begin{tablenotes}
\footnotesize
\item[*]{This column reports “\# prompts that contain direct copy of error messages / \# total prompts.”}
\item[\textdagger]{Live-stream videos were classified as high-reliance if over 40\% of their prompts were method-redundant as defined in §\ref{subsec:roll-the-dice}, and as low-reliance if they conducted substantive code inspection in more than three sessions as described in §\ref{subsec:session-analysis}. Those between these thresholds were labeled medium-reliance.}
\end{tablenotes}
\end{threeparttable}
\end{table}

Reflecting upon the diverse practices, \Otwo{} advocated for a strict definition—vibe coding should be limited to avoid reading diffs or code, and rely on debugging by copy-pasting error messages without inspection, as he believed \textit{``I think that both dilutes the term and gives a false impression of what's possible with responsible AI assisted programming.''} In contrast, \Oseven{} dismissed such boundaries between traditional software engineering and vibe coding, arguing that the best results come from using whatever techniques work. As such, vibe coders expressed mixed feelings about vibe coding. Some criticized it for introducing security risks (\Vtwo, \Vfour, \Vsix, \Oone, \Ofour, \Ofive, \Osix, \Oeight, \Onine, \Oten) or for reducing valuable learning opportunities (\Othree, \Ofour, \Ofive). Others emphasized its advantages, such as broadening access for users with limited technical backgrounds (\Otwo{}, \Ofour{}, \Oeight{}, \Oten{}, \Otwelve{}), accelerating software development (\Othree, \Osix, \Othirteen), and making the process more enjoyable (\Vone, \Vsix, \Osix, \Oten). 

\subsubsection*{\findingrel{finding:vibe-trust}{Vibe coders purposively, unwillingly, or selectively trust AI}}
Across the corpus, vibe coders displayed different modes of trust in AI outputs, ranging from deliberate performance choices to reluctant reliance to more calculated selectivity. Some creators practiced purposive trust (\Vone, \Vfour{}, \Vsix{}), accepting outputs without scrutiny for their audience. \Vone{} described this explicitly as part of “vibe coding”: \cSay[\Vone{}]{We are just full out vibe coding… let’s go ahead and accept all without even reading.} Here, trust was not naïve but staged for effect. 

At the other end of the spectrum were instances of unwilling trust. Coders like \Vfive{} voiced skepticism and frustration yet still deferred most tasks to the AI, admitting, \cSay[\Vfive{}]{I assume the AI knows what it’s doing… I’m getting the same errors... and it’s very frustrating.} Despite these doubts, the same video showed \Vfive{} enabling “YOLO mode” in Claude Code, which allowed terminal commands to run locally without permission—an extreme act of reliance that contrasted sharply with the caution displayed by others. In this sense, trust was not freely given but compelled by circumstance.

More often, coders demonstrated selective trust, calibrating when to rely on AI and when to intervene. Professional programmers selectively treated certain tasks as not worth careful verification. \Vtwo, for instance, remarked, \cSay[\Vtwo{}]{I’ll let the AI generate some design [CSS] that isn’t best practice… I don’t care.} \Vsix{} alternated between acceptance and scrutiny, checking 28 of 54 prompts and verifying outputs for others. At the same time, many videos cautioned against the risks of trusting or executing AI-generated code (\Vtwo, \Vthree{}, \Vfive{}, \Vfour{}, \Othree{}, \Ofour{}, \Ofive{}, \Osix{}), particularly terminal commands (\Vfour{}, \Osix{}, \Oeight{}). As \Oeight{} warned, \cSay[\Oeight{}]{You’re basically giving someone access to your computer to run arbitrary code — in cybersecurity terms, that’s very insecure.} 

Such selective trust was also evident in how coders constructed their prompts. The word choice of prompts mattered to the vibe coders (\Vtwo, \Vfour{}, \Vsix{}, \Oeight{}), while many were unconcerned with typos, transcription errors (\Vone, \Vfour{}, \Vfive{}, \Vseven{}, \Oseven{}, \Otwelve{}), or even in different natural language (\Otwelve{}). Coders assumed the AI could tolerate surface-level messiness but doubted its ability to interpret vague or underspecified intent. \Vtwo{} deliberately inserted the word “properly” into a prompt (\textit{“so that they [as tests] properly pass”}) and explained, \textit{“I don't want [the AI] just going to do assert true.”} \Vsix{} likewise worried aloud, \cSay[\Vsix{}]{I don’t think it was very good instructions.} By contrast, \Vseven{} submitted the garbled request \cSay[\Vseven{}]{use that component in the goals screena dn aslop the user profile screen,} and \Vfour{} sent the prompt with missing information \cSay[\Vfour{}]{I can move around with W[AS]D and I can jump.} As \Oseven{} put it, \cSay[\Oseven{}]{AI is so tolerant of minor grammar and punctuation mistakes that doesn't matter if the transcription's not perfect.} In these cases, trust extended to the AI’s ability to recover intent despite noise.

\subsection{Wrapper \& Tools\label{wrappers}}
\subsubsection*{\findingtools{finding:wrappers}{Vibe coders mapped specific wrappers to tasks, combining them in workflows.}} For example, some advised (\Vfive{}, \Onine{}) or directly built (\Vsix{}, \Oeight{}, \Oten{}) software using online platforms task-specific platform such as VZero, Replit, or Bolt.new, before refining it in local AI code editors like Windsurf or Claude Code. \Othree{} recommended using multiple tools in sequence—for instance, applying a second tool for code review to gain an additional perspective. \Vsix{} endorsed VZero for initial prototyping and praised its frontend aesthetics. Wrappers with built-in Q\&A functions (e.g., Ask Mode in Cursor) were also common, yet several coders (\Vtwo, \Vthree{}, \Vfive{}, \Vsix{}) still switched to ChatGPT when having questions. Vibe coders’ mental models of these wrappers were shaped by feedback from audiences (\Vone, \Vtwo, \Vfive{}), online resources (\Vtwo), or accidental discoveries (\Vone, \Vtwo, \Vsix{}). In one case, audience input led \Vfive{} to adopt the keyword “ULTRATHINK” in Claude Code, exploiting a hidden feature to maximize token budget for deeper reasoning. Or, \Vsix{} accidentally uncovered Windsurf preview feature and exclaimed~\cSay[\Vsix{}]{You can just preview here. It's kind of cool.}

\subsubsection*{\findingtools{finding:opaque-behavior}{Opaque tool behaviors—such as hidden context sent to the model (\Vone, \Vtwo) and the hidden cost of token size (\Vfive{}, \Vsix{})—hindered accurate mental model updates.}} 
When tools are wrapped for model use, additional context (e.g., terminal outputs, or files) may be passed to the model without being visible to the user. This hidden context and system prompt (a predefined prompt used by wrapper) can shape the model’s responses in ways that vibe coders cannot anticipate. For example, due to insufficient documentation of the tool’s capabilities, \Vtwo{} incorrectly assumed the agent could fetch terminal output and issued a debugging request without providing the necessary context, thus failing to debug the error. We, as authors, identified it only by reviewing the GitHub pull request for documenting tool's capability~\cite{githubIssue}.
Likewise, when \Vfive{} noticed that a single task consumed an unexpectedly large number of tokens, they became concerned that the upcoming auto-compaction would reduce work quality.

\subsection{Foundation Model (FM)\label{FM}}
\subsubsection*{\findingtools{finding:models}{Similar to the selection of wrappers, vibe coders purposively selected (all Vs) or combined (\Vsix{}, \Vseven{}) models to achieve their goal.}} Choices were shaped by quantitative factors such as release date (\Vone, \Vtwo{}), budget (\Vfive{}, \Vsix{}, \Ofour{}, \Osix{}), and speed (\Vsix{}), alongside qualitative considerations like trust (\Vfive{}, \Vsix{}), response style (\Vthree{}, \Vfour{}, \Oseven{}, \Oeight{}), and task type (\Vthree{}, \Vsix{}, \Osix{}, \Oseven{}). Some coders also switched models opportunistically to resolve persistent issues (\Oseven{}, \Oeight{}). For instance, \Vsix{} alternated between a free base model for trivial edits (e.g., adding semicolons) and a paid model when greater speed and reliability were desirable, remarking \cSay[\Vsix{}]{[for this task,] I don't trust [the free model]}. \Vthree{} preferred ChatGPT for its tutorial-like replies, noting, \cSay[\Vthree{}]{What I like is it gives you this very simple step-by-step}, while \Vone{} prioritized recency, preferring \cSay[\Vone]{the latest, the greatest}. Similarly, \Oeight{} experimented with model switching to tackle persistent bugs, echoing \Oseven{}’s advice: \cSay[\Oseven{}]{if it's not working, switch models}. Taken together, these practices illustrate a flexible approach to model use, captured in \Othree{}’s suggestion: \cSay[\Othree{}]{do not assume that you'll always use the same agent or the same model. AI space is changing very fast.}

\subsubsection*{\findingaut{finding:unpredictability}{Vibe coders illustrated the unpredictability of FM's behavior while relying on evaluation practices that are themselves unreliable.}} They formed the mental models of the FM’s capacities by making assumptions (\Vone, \Vtwo{}, \Vfour{}, \Vfive{}, \Vsix{}, \Oone{}, \Othree{}, \Ofour{}, \Oseven{}), consulting documentation (\Vtwo{}), or asking the model directly about its capabilities (\Vone, \Vtwo{}, \Vthree{}). For example, \Vtwo{} cross-checked the release dates of a library and the FM in use to assess whether the model could plausibly “know” about a newly introduced feature. Similarly, \Vone{} assumed the model “had an idea” of the game they were working on, thus reducing the information needed in prompts. Such understanding of model capabilities also shaped their artifact choices, including the tech stack (\Oone{}, \Ofour{}, \Oseven{}) and project size (\Othree{}, \Oseven{}). For example, \Oone{} purposively selected a library: \cSay[\Oone{}]{rather than using my preferred tech stack, I decided to [use a library] to build something [faster], especially given that LLMs tend to prefer using it}, though \Ofour{} cautioned that the prevalence of insecure examples in public code for some libraries might lead models to propagate those practices. Meanwhile, \Othree{} emphasized that \cSay[\Othree{}]{the smaller the codebase, the easier it will be for AI to digest it.}, thus advocating microservices architecture.

They noted the model’s inconsistency (\Vsix{}, \Ofour{}, \Ofive{}, \Oeight{}, \Oten{}), its declining performance during long tasks (\Vfour{}, \Vfive{}), and voiced skepticism toward benchmarks (\Vfive{}). Some sought to uncover the model’s reasoning through direct questioning, even though prior work shows such explanations are often plausible but unfaithful~\cite{sarkar_large_2024, madsen_are_2024}. \Vone{} and \Vtwo{} probed hidden properties such as training cutoff dates by asking, \cSay[\Vone{}]{Does your training include experimental features?} Echoing prior literature that model performance may shift over time~\cite{chen_how_2024}, \Vfour{} and \Vfive{} described the system as becoming \cSay[\Vfour{}, \Vfive{}]{dumb} after prolonged debugging. \Vfive{} further rejected formal benchmarks altogether, stating, \cSay[\Vfive{}]{I don't believe benchmarks at all,} and instead relied on hands-on testing through direct interaction.

\subsection{Artifacts\label{artifacts}}
\subsubsection*{\findingrel{finding:artifacts-understanding}{Vibe coders formed mental models of artifacts through prior experience (All Vs), interactions with AI models (\Vone, \Vtwo{}, \Vfive{}), outputs (All Vs), or directly from code (\Vone, \Vtwo{}, \Vthree{}, \Vsix{}).}} 
For those who engage directly the code, comprehending and editing the AI-generated code is similar to working with the code written by others~\cite{maalej_comprehension_2014}. They checked the code (\Vone, \Vtwo{}, \Vthree{}, \Vsix{}), performed refactoring (\Vsix{}, \Vseven{}), printed variables to understand run-time behaviors (\Vtwo{}, \Vsix{}), crafted automated tests (\Vtwo{}), and modified the code manually or with inlined prompts (\Vtwo{}, \Vthree{}, \Vsix{}). However, for those who ``forget the code exist''~\cite{andrejkarpathy} relied on interaction with UI outputs or AI responses. For example, \Vfive{} relied on model to identify web routes they had built and prompted \cSay[\Vfive{}]{What route is my admin page?}, or \Vone{}, when losing track of generated contents, prompted \cSay[\Vone{}]{Where are we with this?}. Some coders even treated the code as “disposable” (\Vone, \Vfour{}, \Otwo{}), wiping out previous work with ease. As \Vone{} explained: \cSay[\Vone{}]{This is the thing about Vibe coding—I have no care about this code. I did not put any of my heart and soul into it.}

\subsubsection*{\findingrel{finding:prompt-shaped-by-understanding}{Vibe coders' mental model around artifacts may affect prompting strategies.}}
Vibe coders who possessed technical knowledge or had reviewed the code could construct prescriptive prompts with specific contexts.
After inspecting the code, for instance, \Vsix{} remarked, \cSay[\Vsix{}]{I have enough information to ask a good question.} and pointed to specific component through prompt: \cSay[\Vsix{}]{The focus-visible box shadow on the advanced options on the left...isn't visible.. can you fix that?} \Vseven{} relied on their understanding of the file structure to supply prompts with specific files or folders as context. 
Many vibe coders emphasize on how expertise could affect the granularity or even correctness of the prompts (\Vsix{}, \Ofour{}, \Osix{}, \Oeight{}, \Onine{}, \Oten{}, \Othirteen{}). For example, \Othirteen{} mentioned \cSay[\Othirteen{}]{you can have AI write a bunch of code [in certain framework] but do you know if you need to use [the framework] for your project?} \Ofour{} discussed the granularity of the prompt, \cSay[\Ofour{}]{[without caring the code,] you don’t go [prompt] create a [specific] class or start refactoring.}

Vibe coders with limited technical backgrounds, or who never examined the code, tended to provide instructions and sometimes introduced errors. \Vfour{}, for example, issued prompts containing only one library name during setup and later reflected: \cSay[\Vfour{}]{I just am not prompting correctly… I don’t know enough about 3D game engines to tell it how to.} \Oeight{} revealed a mismatch in their understanding of the project, noting: \cSay[\Oeight{}]{I thought I was making an [mobile] app, apparently, this is a website.} \Vfive{} when facing a race condition, suggested adding delays to database writes rather than using locks to manage shared resources. To mitigate potential mistakes, some vibe coders sought outside expertise—\Oten{}, for instance, consulted professional developers to identify security issues and emphasized the value of deliberation with AI before implementation: \cSay[\Oten{}]{talking each problem through with the AI before you start telling it what to do is always going to be better.} Others recommended “learning with AI” (\Vtwo{}, \Vthree{}, \Vsix{}, \Oseven{}, \Oeight{}, \Onine{}, \Oten{}). As \Onine{} explained: \cSay[\Onine{}]{[ask AI] if you're learning about the frameworks that are being used while you're vibe coding, this is going to significantly improve your skills as a vibe coder}.

\subsection{Prompts \& Contexts \label{prompts}}
\subsubsection*{\findingtools{finding:prompting-strategies}{Vibe coders developed and employed a range of strategies to execute requests, to explore, or to understand the current situation.}}
To prompt models to fulfill requests, vibe coder employed a variety of strategies, including those previously described in literature, such as few-shot prompting (\Vtwo{}) and persona patterns (\Vfour{}, \Vfive{}, \Oeight{}). Some vibe coders developed their own strategies, such as inserting emotional cues (\Vfour{}, \Vfive{}, \Oeight{}). For instance, \Vfour{} experimented with various approaches, instructing the model with directives like  \textit{``you are a 3D game engineer''}., or purposefully shifting emotional tone between statements such as \textit{``I didn’t ask you to''} and \textit{``I apologize for being angry''}. Prompts were also used to explore and guide implementation decisions (\Vone, \Vtwo{}, \Vthree{}, \Vsix{}, \Oten{}). For instance, \Vone{} asked: \textit{``Help me think through how the UI will work.''} Similarly, \Vthree{} inquired: \textit{“I’m about to create a project with Rails… what should I do next?”} Finally, prompts sometimes were used to support sense-making, such as asking the model to describe its reasoning (\Vone, \Vtwo{}) or asking the model to provide explanations to help with understanding the generated code or outputs (\Vone, \Vtwo{}, \Vfive{}, \Vsix{}). Examples include \Vone’s expression of skepticism when seeing the outputs: \textit{``Is that rooted with truth?''} or \Vtwo{}’s use of a screenshot to ask: \textit{“Explain this code.”}

\subsubsection*{\findingtools{finding:context-manage}{Vibe coders actively manage the context to avoid over-indexing (\Vone, \Vfour{}, \Vsix{}, \Othree{}, \Oseven{}, \Oten{}) and reduce context window used (\Vfive{}, \Oten{}).}} Vibe coders sometimes start fresh sessions (\Vone, \Vfive{}, \Vsix{}, \Othree{}, \Oseven{}, \Oten{}) or remove lingering artifacts from prior work (\Vone, \Vfour{}) to avoid polluting the future responses. For example, while aiming for switching programming languages to Python, residual JavaScript files caused persistent JS generation, so the \Vone{} cleared the codebase and instructed the model to ignore deleted code. \Vsix{} cleared chat histories for each new feature to prevent over-indexing on prior output. \Oten{} forked the projects as for VZero as \cSay[\Oten{}]{the context window got too long and the AI was confused, this is when I would fork.} In hope to generate better results, they asked AI to fetch context (\Vfive{}) like \cSay[\Vfive{}]{start by reading my entire codebase} or manually add relevant context—via web search tools (\Vtwo{}), images (\Vfour{}, \Vfive{}, \Vseven{}, \Onine{}), point to local files (\Vone, \Vsix{}, \Vseven{}), and terminal (\Vone, \Vseven{}), or the Model Context Protocol (MCP) (\Vfive{}, \Othree{}). Inefficient copy-and-paste remained widespread (all Vs except \Vseven{}), though \Osix{} and \Oseven{} anticipated it would soon be automated. For example, \Vfour{} provided a reference image to guide 3D shader patterns; \Vfive{} loaded a Shopify MCP so Claude Code could use the provided APIs; and \Othree{} recommended loading a memory graph with MCP and regularly refreshing chat histories.

\subsubsection*{\findingtools{finding:tracking-shared}{Keeping track of what has been shared with models has emerged as challenges for vibe coders.}} Vibe coders sometimes reminded the AI of prior prompts (\Vone, \Vfour{}, \Vsix{}) or revisited prior chat histories (\Vtwo{}, \Vfive{}) to determine whether specific questions or requirements had already been provided. For example, \Vfive{} searched their ChatGPT history to locate previously answered questions, thereby avoiding redundant prompts and reusing prior context. In addition, memory lapses occurred: vibe coders (\Vfour{}, \Vfive{}) expressed uncertainty about whether certain instructions had already been provided, and \Vfour{} initially requested a \emph{30$\times$30$\times$30} tile grid (a 3D structure), but later asked why a \emph{20$\times$20$\times$20} grid was missing.
Some vibe coders also expect AI to have human capabilities~\cite{zamfirescu-pereira_why_2023}, and felt frustrated by the model’s limited retention of prior requests (\Vone, \Vfour{}, \Vfive{}), like \Vone{} noted, \textit{``my job is to remind it of the things I’ve told it before''}. Some suggested or implemented a more systematic strategy by creating a \textit{progress.md} (\Vone) or instruction file (\Oseven{}, \Oten{}) or using external memory graph (\Othree{}) to externalize and track sent content across sessions.

\subsection{Requirements \& Design\label{requirement}}
\subsubsection*{\findingdes{finding:reqs-upfront-emergent}{While some vibe coders created (\Vone, \Vthree{}) and advocated (\Othree{}, \Onine{}, \Oeleven{}) detailed requirements prior to vibe coding, many formulated new requirements in response to outputs.}} For example, \Vthree{} produced a detailed product requirements document (PRD) before vibe coding and copied the entire document into the model when creating the software. Similarly, \Oeleven{} emphasized the value of a detailed PRD, noting: \cSay[\Oeleven{}]{[with PRD], the less engagement we need to correct during the process.} At the same time, many vibe coders (\Vone, \Vtwo{}, \Vfour{}, \Vfive{}, \Vsix{}, \Vseven{}, \Oten{}) also developed requirements during the process in ways resembling exploratory programming~\cite{beth_kery_exploring_2017}. For instance, \Vone{} co-authored detailed requirements with the AI model prior to coding, but upon seeing outputs, formulated new requirements such as revising the user interface or adding new game mechanics. \Vseven{}, when refining a feature, first offered a loosely defined prompt—\cSay[\Vseven{}]{there should be an effect to this, like a nice loading circle or something}—before incrementally refining it across later prompts. Likewise, \Oten{} explained: \cSay[\Oten{}]{I know people said to plan out a perfect PRD. But… there’s so many things that you don’t think about until you have to actually have the AI code them.}

\subsubsection*{\findingdes{finding:fixation}{Vibe coders typically accepted or refined the first generated outputs rather than seeking alternatives, signaling design fixation. (All Vs)}} 
After issuing loosely defined prompts, such as \Vsix{} prompting, \cSay[\Vsix{}]{Build a modern website UI} or \Vone{} prompting, \cSay[\Vone{}]{Make an app that is a better version of the UI for [the game]} (and as seen with \Vseven{} above), coders typically iterated on the first result instead of prompting the model for alternative directions. While this pattern dominated live-streamed sessions, \Oseven{} suggested a contrasting approach: running the same prompt across two different tools and then, as they explained, \cSay[\Oseven{}]{pick[ing] which one I like better.}

\subsection{Implementation \& Debugging\label{implementation}}
\subsubsection*{\findingaut{finding:stochastic-debug}{Debugging or feature refinement---expected deterministic---can feel stochastic, like \textit{``throwing a dice''} (\Ofour{}, \Oseven{}).}} Unlike traditional debugging strategies, such as fault localization~\cite{jones_visualization_2002}, or hypothesis formation~\cite{alaboudi_hypothesizer_2023}, some vibe coders with high AI-reliance (\Vfour{}, \Vfive{}) resolved their errors completely by repeating requests with the same or derived error messages without providing any other context or hypothesis. \Vfive{} sent 31 consecutive prompts containing only error messages throughout an hour, hoping Claude Code would eventually fix the issue and stated \textit{``This is what I'm jealous about senior developers. They see kwargs and they would say, oh I know what that means, and I will be what is kwargs?''}. In contrast, \Vtwo{} adapted when AI-assisted debugging failed—reading the code directly and manually resolving the problems: \textit{``Sometimes it overvibes and you just got to do it [real coding].''} Similarly, sometimes a feature refinement request might alter prior working part of the software (\Vfour{}, \Vfive{}, \Vsix{}, \Ofive{}, \Osix{}, \Oseven{}, \Onine{}). \Vfour{} prompted, \textit{``you just made this game worse''} after adding features, noting that previously functional parts of the software broke in the process. \Oseven{} described the AI’s stochastic changes as a \textit{``vision quest,''}, while \Ofour{} rejected this framing, highlighting the differences between programming and image generation: \textit{“[With images], you can just reroll, since an entire class of outputs can fit the problem. But in programming, one pixel off is a bug.”}

\subsubsection*{\findingaut{finding:mitigations}{Vibe coders used undo, version control, small changes, automated tests, and linters to rein in stochastic changes}} As agents' edits could affect unrelated code, coders (\Vfive{}, \Vsix{}, \Osix{}, \Oten{}) sought undo support and version control to restore to a working copy (\Vone, \Vsix{}, \Vseven{}, \Othree{}, \Osix{}, \Oseven{}, \Onine{}, \Oten{}). As \Oseven{} explained: \cSay[\Oseven{}]{[With version control] don’t be afraid to git reset -{}-{}hard HEAD… and roll the dice again.} Similarly, \Vone{} remarked: \cSay[\Vone{}]{Let's do my only job which is to push.} Coders also stressed the value of making small, incremental edits (\Vone, \Vfour{}, \Vfive{}, \Vsix{}, \Oone{}, \Othree{}, \Oseven{}, \Oeight{}, \Onine{}). For instance, \Onine{} advised: \cSay[\Oone{}]{[breaking] up larger features into smaller bite-sized tasks} To safeguard generated results, vibe coders relied on automated tests (\Vtwo{}, \Othree{}, \Osix{}, \Oseven{}) and typed languages for stronger linting support (\Vsix{}). \Oseven{} even advocated test-driven development: \cSay[\Oseven{}]{Handcraft the test cases first… [then] LLMs can freely generate the code… and if I see those green flags the job is done.}

\subsubsection*{\findingtools{finding:wait-frustration}{Waiting for the next output frustrated many (\Vone, \Vtwo{}, \Vfour{}, \Vfive{}, \Vsix{}, \Vseven{}, \Oone{}, \Oseven{}, \Oeight{}); some mitigated this by multitasking (\Vtwo{}, \Vfive{}, \Vsix{}, \Oone{}, \Oeight{}) or sending prompts in parallel (\Vfour{}, \Vseven{}, \Oseven{}).}} For live-streamers, wait time was often used to engage with their audience, yet frustration persisted. As \Vone{} described: \cSay[\Vone{}]{Vibes are kind of like waiting for scroll bars to go across.} Similarly, \Vfour{} noted: \cSay[\Vfour{}]{You have to have patience.} \Oeight{} admitted, \textit{“In between prompts, \ldots I got into a bad habit of scrolling on [videos].”} Vibe coders also shared strategies for coping: \Vtwo{} suggested playing a quick game, \Othree{} recommended verifying generated outputs in parallel, while some suggested running (\Vfour{}, \Oseven{}), or sending multiple prompts (\Vseven{}) in parallel within the same codebase.

\subsection{Evaluation\label{evaluation}}
\subsubsection*{\findingrel{finding:evaluation-variation}{Vibe coders perform evaluations on multiple levels: code, outputs, or responses. Some delegated the evaluation to AI}}
Vibe coders who exhibited low AI-reliance or who had advanced expertise conducted manual testing or co-authored automated tests with AI. \Vsix{} performed manual checks akin to combinatorial testing~\cite{nie_survey_2011} by exercising combinations of available UI options after adding a feature, and raised concerns with the AI: \cSay[\Vsix{}]{If we allow users to add any header they want, is that a security vulnerability?} When no UI was available, \Vtwo{} asked the AI to generate automated tests to isolate functions but did not accept them at face value, noting: \cSay[\Vtwo{}]{I don't think that's quite right—it's saying all the tests are passing, but it probably wrote the tests [so it can pass].} \Vtwo{} then inspected runtime values via prints and brainstormed additional edge cases with the AI, even spotting a bug. In other cases, high AI–reliance coders with irrelevant background conducted evaluation on model responses and outputs. For example, \Vfive{} requested summaries and assessments of terminal output and also prioritize issues: \cSay[\Vfive{}]{Is this error important? Do we need to fix it?} \Vfour{} evaluated the generated results through model responses, for example, after reading the model’s suggested fix, they expressed skepticism: \cSay[\Vfour{}]{Why would that be the issue?}


\section{Additional Analysis}

Building on our results, we investigate how developers’ reliance on AI shapes their vibe coding practices—particularly their time allocation across activities and prompting strategies. We conducted an exploratory mixed-methods study to extend grounded theory findings by revealing behavioral patterns not fully captured through coding alone. Our analysis comprises two components: activity segmentation, and a prompt similarity (\textit{``rolling the dice''}) analysis. 
These exploratory analyses served as deepening understanding for prior qualitative analysis rather than generalization of all vibe coding practices. Below, we detailed the process for each analysis.
\subsection{Sessions Analysis\label{subsec:session-analysis}}
\subsubsection{Motivation} This analysis is motivated by two findings: (1) differences in the practices of vibe coding affect time allocation~\ref{finding:vibe-meaning}, and (2) frustration toward the wait time~\ref{finding:wait-frustration}.  
\begin{figure}
    \centering
    \includegraphics[width=1\linewidth]{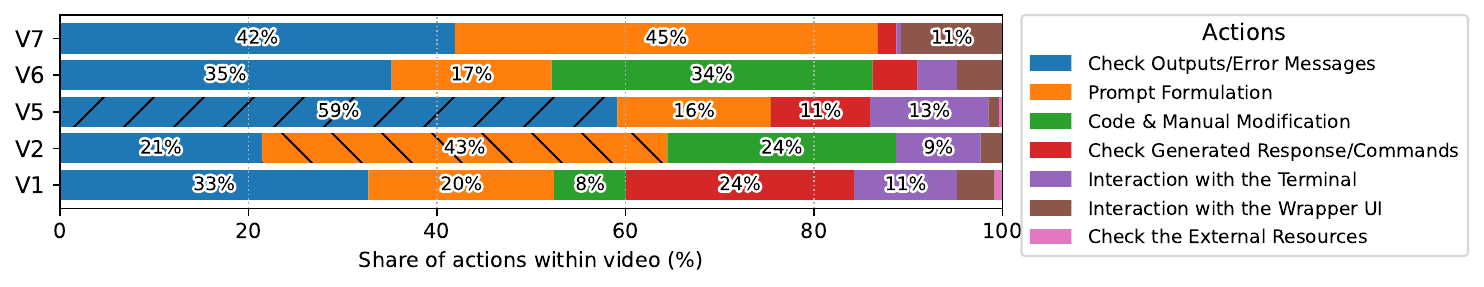}
    \caption{Time allocations for each live-streamed video.}
    \label{fig:time-allocation}
\end{figure}
\subsubsection{Method} To this end, we adopted a method similar to that of Alaboudi and LaToza~\cite{alaboudi_what_2023, alaboudi_exploratory_2021}. After familiarizing ourselves with the collected live-stream videos, four authors met to identify an initial set of activities of interest. We then divided the five live-stream videos among the authors, excluding V3 and V4 as they were edited, each annotating the first five minutes of their assigned videos. Inter-rater reliability reached a substantial level, with a Fleiss’ Kappa of 0.7544. \begin{wrapfigure}{r}{0.4\textwidth}
  \vspace{-\baselineskip} 
  \centering
  \includegraphics[width=\linewidth]{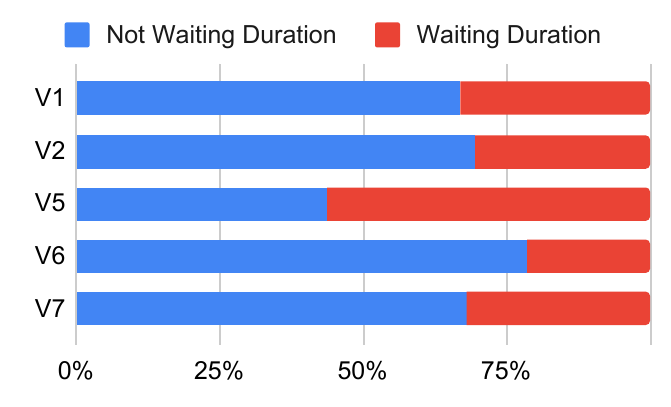}
  \caption{Waiting time as a share of total session duration}
  \label{fig:wa}
\end{wrapfigure} We then reviewed discrepancies collaboratively and refined the definitions of each code. This process enabled us to evaluate the applicability of the initial activity segments and to assess whether the definitions of “start action” and “end action” could be consistently applied across sessions. Following this, the videos were further divided for full annotation by each author, during which the codebook of segment definitions was iteratively refined. The codebook served as the basis for subsequent annotation aimed at supporting the additional quantitative analysis. Using the agreed-upon codebook, each author annotated their assigned videos and obtained 2,439 sessions. We then computed descriptive statistics to summarize the duration of activities across videos. Additionally, we also calculate the waiting time for the generation. Waiting time is the interval between sending a prompt and the author checking the response, ending when generation finishes (if continuously monitoring), required actions are completed, or the vibe coders returns from another task. All annotations could be found in replication package~\cite{Dryad:2jm63xt31}.

\subsubsection{Result} We excluded irrelevant activities, such as audience interaction, and presented the results as a 100\% stacked bar chart in Figure~\ref{fig:time-allocation}. Time allocation patterns highlight how coders experienced vibe coding, extending the contrasts noted in~\ref{finding:vibe-meaning}. For example, V6 and V2—both technically skilled and accustomed to reading code—spent much of their sessions inspecting generated code rather than relying on model explanations. In contrast, V5 and V7 focused more on reviewing outputs and responses without engaging directly with the code. Time spent checking external resources was limited (<5\%), compared to a previously reported median of 15\% for information seeking on the web~\cite{xia_what_2017}. With respect to waiting time—where coders expressed frustration (~\ref{finding:wait-frustration})—we observed in Figure~\ref{fig:wa} that over 20\% of total session time across all livestreams was spent waiting for generation. For V5, this proportion even exceeded half of the entire session.

\subsection{Rolling the Dice Analysis\label{subsec:roll-the-dice}}
\subsubsection{Motivation} 
Our Finding~\ref{finding:vibe-meaning} reveals variation in how vibe coders define and enact vibe coding, 
particularly in their engagement with code. As shown in Table~\ref{tab:different_behavior}, some coders (e.g., V2, V6) inspected diffs and refined prompts, while others (e.g., V5, V4) never read code and often repeated requests or copied error messages. These contrasts motivate our analysis of \textit{rolling the dice}, where developers reissue the prompts with similar intent without adding new information.
\subsubsection{Method}We developed the initial code set iteratively through group discussion. The first four authors then collaboratively applied these codes to all prompts in a single coding session. We documented code definitions and identified edge cases throughout the process; both are included in the replication package. Each prompt was segmented into units and labeled by intent, defined as the underlying objective of the prompt (e.g., fixing an error or adjusting a component’s height). Prompts sharing the same intent were subsequently analyzed for method-level variation. A prompt was coded as method-redundant (or rolling the dice) if it repeated a prior intent without introducing new information, such as simple rewording, copying an identical error message, or rejecting an output without further explanation. Prompts that introduced new details or hypotheses were not considered rolling the dice. When a prompt contained multiple intents, it was coded as intent-partially-redundant if any of its intents overlapped with those in the immediately preceding prompt.

\subsubsection{Result}
\begin{wrapfigure}{r}{0.5\textwidth}
  \vspace{-\baselineskip} 
  \centering
  \includegraphics[width=\linewidth]{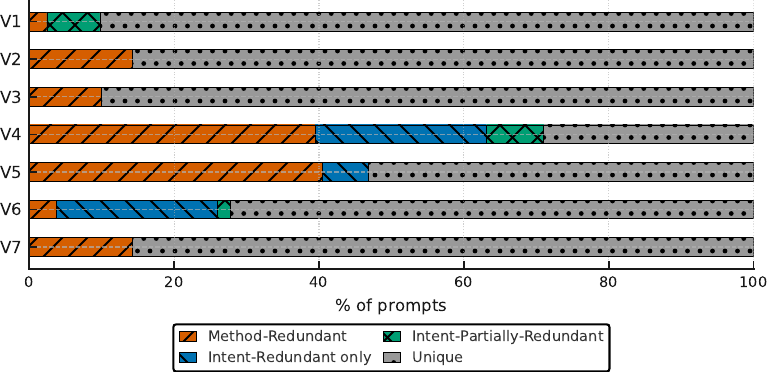}
  \caption{Normalized stacked bar chart of prompt redundancy for vibe coders V1–V7. Bars show proportions of method-redundant, intent-redundant, intent-partially-redundant, and unique prompts}
  \label{fig:roll-the-dice}
\end{wrapfigure}
\textbf{Vibe coders with less relevant technical background and high reliance on AI produced a greater share of redundant prompts.} 
In Figure~\ref{fig:roll-the-dice}, V5 and V4 devoted nearly 40\% of their prompts 
to \textit{rolling-the-dice}, often repeating the same 
requests or reposting error messages without elaboration or new details. 
In contrast, coders who more frequently inspected code
(e.g., V1, V2, V3, V6) or drew on relevant backgrounds (e.g., V7) issued substantially fewer intent- and method-redundant prompts, typically under 20\%. 
V6 provides a notable case: while reissuing the same intents, they avoid \textit{method-redundancy} by adding hypotheses or technical specifics, thereby varying the approach across prompts. This results echoed prior Finding~\ref{finding:prompt-shaped-by-understanding}, where mental model around artifacts might shape the prompting strategies.
\section{Discussion}
In the discussion, we connect our findings on vibe coding to prior software engineering research and discuss the future for researchers and practitioners (§6.1) and outline actionable takeaways (§6.2) for tool designers, educators.

\subsection{Vibe Coding: Rolling the Dice to Build Software}
\subsubsection*{Constructing Opportunistic Workflow}
Vibe coders did not constrain themselves to a single setup, but opportunistically assembled toolchains from multiple models and wrappers (\ref{finding:wrappers}, \ref{finding:models}), aiming for higher quality results or reduced costs. This pattern aligns with Meem et al.~\cite{meemWhyChatGPTPractitioners2025}, who found that developers often use ChatGPT alongside AI-powered editors. Such workflows illustrate a broader shift from writing and maintaining code toward obtaining usable outputs—for example, V6 first leveraged VZero to generate an aesthetic web UI before transferring the project to Windsurf. In this sense, vibe coders were largely indifferent to code structure (\ref{finding:artifacts-understanding}), treating it as a disposable intermediary and prioritizing ``generated results'' over well-formedness. This orientation stands in contrast to the philosophy of IDEs, which treat code as a first-class entity, structured hierarchically and reinforced through features such as syntax checking and code navigation~\cite{teitelbaum_cornell_1981}, and tools that emphasizes code comprehension~\cite{bragdon_code_2010}. Further work is needed to assess whether these practices remain viable in the context of large-scale software maintenance and what forms of tool support could make them feasible.

\subsubsection*{Reining in Stochastic Code Change.} 
Our findings align with recent work on prompt engineering, where developers struggled with the stochastic behavior of FMs~\cite{liang_prompts_2025} and adapted their prompting strategies largely through experience, often in unpredictable ways~\cite{khojahPromptProgramming2025, liang_prompts_2025, tafreshipour_prompting_2025}. Our findings also echo challenges documented in earlier studies of AI-assisted programming tools such as GitHub Copilot, before the rise of agentic workflows: developers reported difficulties in crafting effective prompts for intended outcome~\cite{zamfirescu-pereira_why_2023, liang_large-scale_2024} and in auditing generated outputs~\cite{gordon_co-audit_2023, barke_grounded_2023}. Much of this prior work characterizes \emph{snippet-level} scope. By contrast, vibe coding shifts users away from direct code manipulation, issuing prompts that can trigger repository-wide changes. Once users disengage from code, the stochastic behavior of AI \ref{finding:unpredictability} intrudes on what would otherwise be deterministic code changes \ref{finding:stochastic-debug}, with effects that propagate across debugging, feature refinement, and even evaluation decisions \ref{finding:evaluation-variation}. To mitigate this uncertainty, vibe coders attempted to narrow the scope of changes and relied on timely version control and other safeguards \ref{finding:mitigations}, making these practices essential guardrails for reining in uncertainty at scale. Future SE research could examine ways to safeguard or communicate the stochastic code change to users, such as leveraging the line of work in local code change management and visualization tools~\cite{kery_variolite_2017, yoon_visualization_2013, yoon_supporting_2015}, or leveraging modular-level regression testing to prevent unwanted changes~\cite{al-refai_package_2025, chen_testtube_1994}.

\subsubsection*{Exploring, yet Not Enough}
Vibe coding exemplifies exploratory programming (\ref{finding:reqs-upfront-emergent}): vibe coders iteratively refine requirements in response to outputs, prioritizing rapid experimentation over long-term maintainability. Prior work shows that high-level languages (e.g., JavaScript, Python) enable such iteration by reducing edit distance and offering expressive abstractions~\cite{beth_kery_exploring_2017, sandberg_smalltalk_1988, kery_variolite_2017}. Vibe coding pushes this further by shifting the primary abstraction from code to conversation with AI agents—lowering the cost not just of code edits but of specification changes. As specification and implementation blur, requirements work shifts from documents to executable trials: rather than drafting detailed requirements or user stories, project managers or requirements engineers could “vibe code” to explore and test candidate requirements directly, before handing off prompts, traces, or prototypes for engineering hardening. 

While our findings indicate exploratory behavior, coders often anchor on the first generated outputs and become largely constrained by them, using subsequent prompts primarily to refine the first output. This pattern signals potential design fixation~\ref{finding:fixation}~\cite{jansson_design_1991}. 
Such finding corroborate work by Liang et al.~\cite{liang_large-scale_2024} from the context of early AI coding assistants without agents, who observed that developers valued acceleration mode—where the AI fulfills their intentions—more than exploration mode, where the AI suggests alternative options. 
They also align with Wadinambiarachchi et al.~\cite{wadinambiarachchi_effects_2024} from the context of AI image generation, who found that designers ideating with AI-generated images exhibited stronger fixation and produced fewer, less original ideas compared to baseline groups. 
In our context of vibe coding, such fixation can have large downstream effects. 
For example, an initial misinformed design decision of the software architecture could cascade to the entire software development process, leading to piles of technical debt~\cite{SOLIMAN2021106669}. Future research could investigate how to mitigated AI-induced design fixation, for example, through visualization for structured design space~\cite{suh_luminate_2024}.
\subsubsection*{Evolving Definitions of Vibe Coding}
The term vibe coding, as discussed in (\ref{finding:vibe-meaning}), encompasses a diverse range of practices. Our study provides a snapshot of current approaches without attempting to prescribe or delimit what should formally qualify as vibe coding. Because it lowers the barrier to software creation, practitioners bring widely varying backgrounds, expertise levels, and personal definitions of the practice—differences that shape their behaviors and mental models. 
Our finding in prompting strategies echo a finding from Geng et al.~\cite{gengexploringstudentaiinteractionsvibe2025}, who observed that more experienced students tended to provide prescriptive prompts and intervene when needed, as well as Zamfirescu et al.~\cite{zamfirescu-pereira_why_2023} where they found non-experts often struggle finding the right instructions and repeating similar requests with ``retries'' without reasoning the potential prompt design issues.  
While we highlight how differences in mental models, reliance on AI, and background influence prompting strategies (\ref{finding:prompting-strategies}), evaluation practices (\ref{finding:evaluation-variation}) and trust (\ref{finding:vibe-trust}), future research is needed to continue reexamining this evolving construct.

\subsection{Takeaways\label{takeaways}}
\subsubsection{For Tool Designers,} our findings reveal that vibe coders sometimes lost track of prior prompts and evolving goals~\ref{finding:tracking-shared}. 
Unlike prior challenges in software engineering, where design decisions often remain tacit and undocumented~\cite{ryan_acquiring_2013}, future tool could mitigate through collecting, and summarizing the sent prompts within IDE with insights from research in decision tracking ~\cite{mangano_design_2012, horvath_meta-manager_2024}. 

We also observed variation in how expertise affects how vibe coders engaged with artifacts and evaluated outputs (\ref{finding:evaluation-variation}, \ref{finding:prompt-shaped-by-understanding}), such as reiterating repetitive prompts or having limited evaluation strategies. Task decomposition also remains a particular challenge: while coders recommended breaking work into small chunks to avoid unintended changes (\ref{finding:mitigations}), they still needed enough knowledge to judge what constitutes “small.” This includes anticipating how a given prompt might affect the codebase and evaluating which components it touches. For example, one of the longest wait times we observed—27 minutes in session V5—stemmed from a single prompt that ranged from swapping databases to revamping UI/UX. To accommodate vibe coders from diverse background and engagement with code, systems could adapt responses—providing scaffolding for novices~\cite{ma_dbox_2025, tankelevitch_metacognitive_2024} while remaining unobtrusive for experienced users to accelerate their well-formed intents~\cite{barke_grounded_2023}.

For mitigating unpleasant wait times (\ref{finding:wait-frustration}), vibe coders attempt to cope by submitting prompts in parallel. Such practices, yet, introduce potential conflicts—as seen in \Vseven{}, where file-level locks and manual audits were required. Future work should further investigate conflict resolution in multi-agent systems and approaches for reducing developers’ auditing burden. Additionally, tool designers can better support developers in sustaining flow~\cite{ritonummi_flow_2023} by providing code-change explanations that help bridge abstraction gaps~\cite{liu_what_2023, subramonyam_bridging_2024} and by proactively proposing questions or challenging incorrect instructions to prevent unproductive paths~\cite{sarkar_ai_2024}. Copy-and-pasting error messages or terminal outputs was also cumbersome; tools that import these automatically could reduce this burden (\ref{finding:context-manage}). Finally, vibe coders struggled with hidden behaviors of wrappers, such as undisclosed context, system prompts, or token costs (\ref{finding:opaque-behavior}). Improving transparency in these areas could help vibe coders form more accurate mental models and calibrate their trust accordingly~\cite{wang_investigating_2024, al-ani_globally_2013}.

\subsubsection{For Educators,} Our study shows that vibe coders with even partial knowledge (e.g., \Vone{}, \Vseven{}) for artifacts produced more detailed prompts and evaluated outputs more effectively, whereas those relying heavily on AI (e.g., V4, V5) struggled to provide detail or assess results (\ref{finding:evaluation-variation}, \ref{finding:prompt-shaped-by-understanding}). As Miyake and Norman note, ``To ask a question, one must know enough to know what is not known''~\cite{miyake_ask_1979}. Although some advocate learning directly with AI to mitigate errors (\ref{finding:prompt-shaped-by-understanding}), this approach risks leaving learners unable to ask the right questions or detect suboptimal outputs (e.g., O8 did not realize they were building a web rather than a mobile app in \ref{finding:prompt-shaped-by-understanding}). We argue that deductive education remains essential: by first teaching general principles and conceptual frameworks (e.g., synchronization using locks for shared resources) and then guiding students to apply them in practice (e.g., handling multiple concurrent requests when building applications, as contrasted with V5 in \ref{finding:prompt-shaped-by-understanding}), educators provide learners with the evaluative lens needed to critically assess AI outputs (e.g., V6 raised concerns in \ref{finding:evaluation-variation}). Additionally, to prepare students for working with LLMs—rather than humanizing their capabilities~\cite{zamfirescu-pereira_why_2023}—educators should cultivate basic data-science literacy regarding AI capabilities, limitations, and evaluation strategies (\ref{finding:unpredictability}).

\section{Conclusion}
This study sheds light on how vibe coders build software by “rolling the dice” with AI-powered tools. Through analysis of livestreams and opinion videos, we showed how practitioners construct opportunistic workflows, improvise requirements and design, and develop diverse strategies shaped by their mental models. These practices reveal the creative potential of vibe coding to broaden participation in software creation and support rapid experimentation. At the same time, the reliance on stochastic AI outputs introduces challenges—such as fragile evaluation practices, early anchoring on initial outputs, and debugging that feels probabilistic rather than deterministic. Recognizing both the opportunities and tensions, our findings point to important directions for tool design and education to harness the benefits of vibe coding while addressing its limitations.
\section{Data Availability}
The dataset consists of public, searchable YouTube videos and contextual information cross-referenced from LinkedIn and individual's public website. The YouTube links will be visible during peer review for transparency but anonymized in the final publication to minimize risks of unwanted exposure, according to ethical recommendation from Townsend and Wallace ~\cite{townsend2016social}. We provide a replication package~\cite{Dryad:2jm63xt31} with our codebook, coded excerpts, analysis scripts. For individual information, we include a summary sentence per individual rather than raw profiles. 
\section{Acknowledgments}
We thank the content creators whose publicly available videos made this research possible,
and the anonymous reviewers for their constructive feedback.
Yi-Hung Chou was supported in part by the Taiwan Government Scholarship to Study Abroad.
The views and opinions expressed in this work are solely those of the author and do not necessarily reflect the views, policies, or positions of the author’s employers.
\bibliographystyle{ACM-Reference-Format}
\bibliography{references, vibecoding}


\end{document}